\documentclass{kapproc} % Computer Modern font calls
\usepackage{procps} 
\usepackage{amssymb}
\usepackage[dvips]{graphicx}

\upperandlowercase

\nochapequationnumber % will result in equation numbers that are (1)

\let\footnote\savefootnote

\let\footnotetext\savefootnotetext

\normallatexbib %will produce bibliography entries as shown in the
\begin{document}

\articletitle{QCD hard scattering results from PHENIX at RHIC
%\footnote{NATO-ASI %Advanced Studies Institute
%"Structure and Dynamics of Elementary Matter", Kemer, Turkey, Sept. 22
% -- Oct. 2, 2003}
}
%\articlesubtitle{This is an Article Subtitle}

\author{David d'Enterria for the PHENIX Collaboration}
\affil{Nevis Laboratories, Columbia University\\
Irvington, NY 10533, and New York, NY 10027, USA}
\email{denterria@nevis.columbia.edu}

\begin{abstract}
Data on hadron production at high transverse momentum ($p_{T}>$ 2 GeV/$c$) 
in p+p, d+Au, and Au+Au collisions at $\sqrt{s_{_{NN}}}$ = 200~GeV 
from the PHENIX experiment at RHIC are reviewed. The single inclusive 
spectrum of light %-flavored 
hadrons produced in central Au+Au reactions shows significant differences 
compared to p+p and d+Au collisions, and provides interesting information on the 
properties of the underlying QCD medium present in heavy-ion reactions at collider energies.
\end{abstract}


\section{Introduction}
The fundamental degrees of freedom of the theory of the strong interaction
(Quantum Chromodynamics, QCD) are colored quarks (fermions) and gluons 
(gauge bosons). In the real world, however, quarks and gluons 
are not observed free but confined in color singlet states (hadrons).
In addition, the observed effective {\it constituent} mass of the quarks 
(${\bar m_{u,d}}\approx 300$ MeV for the lightest $u$ and $d$ quarks) is much larger 
than the bare {\it current} mass of the QCD Lagrangian 
($m_{u,d}\approx 2 - 8$ MeV). %As a matter of fact, t
Theoretically understanding from 
first principles these two basic properties of QCD: confinement and chiral symmetry 
breaking, remains in fact one of the most important problems in fundamental physics~\cite{millenium_prizes}.
Lattice calculations of QCD in the bulk~\cite{latt} predict that above an energy density of 
$\epsilon_{crit}\approx$ 0.7 $\pm$ 0.3 GeV/fm$^3$, the (long range part of the) 
strong color potential becomes screened and, as a result, quarks and gluons 
can move freely with their bare mass within the deconfined medium. Experimentally 
accessing and studying the properties of this ``Quark-Gluon Plasma'' (QGP) phase 
is one of the main goals of the physics program of the PHENIX experiment at the Relativistic 
Heavy-Ion Collider (RHIC) at BNL. By colliding heavy nuclei at high energies one expects to produce
a thermalized system of quarks and gluons with energy densities above $\epsilon_{crit}$,
albeit for very short time scales, $\mathcal{O}(10^{-23}$ s) and small volumes, 
$\mathcal{O}(10^{2}$ fm$^{3})$.
In the period 2000-2003, PHENIX collected data from Au+Au, d+Au and p+p collisions
at $\sqrt{s_{_{NN}}}$ = (130)200~GeV. The most interesting phenomena observed
so far in the data are in the high $p_T$ ($p_T\gtrsim$ 2 GeV/$c$) 
sector where the production of hadrons in (non-peripheral) Au+Au collisions shows 
substantial differences compared to more elementary reactions either in 
free space (p+p, $e^+e^-$) or in a ``cold'' nuclear matter environment (d+Au).
%Events with large momentum transfer in these reactions are the result of violent short-distance collisions. 
%Such {\it hard scattering} processes posses very interesting properties:
Hard scattering processes are an excellent experimental probe in heavy-ion collisions
inasmuch as: %they posses the following interesting properties:
(i) they are the result of violent (large $Q^2$) short-distance (parton-parton) 
collisions in the first instants of the reaction ($\tau\sim 1/p_{T}\lesssim$ 0.1 fm/$c$) 
and, as such, direct probes of the {\bf partonic} phase(s) of the reaction;
(ii) in the absence of medium effects, their cross section in A+A
reactions is expected to {\bf scale simply} with that measured in {\bf p+p} 
collisions times the number of %available point sources
scattering centers ($A^2$, see discussion below); and
(iii) their production yields can be {\bf theoretically calculable} in QCD via standard 
perturbative (collinear factorization plus Glauber multiscattering~\cite{pQCD}) or 
via ``classical-field'' (``Color-Glass-Condensate'' approach~\cite{CGC}) methods.

\section{The PHENIX experiment}
The PHENIX detector~\cite{nim} is specifically designed to measure hard %(``penetrating'') 
QCD probes such as high $p_T$ hadrons, direct photon radiation, lepton pairs, and heavy flavour production.
PHENIX achieves good mass and PID resolution, and small granularity by combining 13 detector 
subsystems ($\sim$350,000 channels) divided into: (i) 2 central arm spectrometers for
electron, photon and hadron measurement at mid-rapidity ($|\eta|<0.38$, $\Delta\phi=\pi/2$); 
(ii) 2 forward-backward ($|\eta|$ = 1.15 - 2.25, $\Delta\phi = 2\pi$) spectrometers for 
muon detection; and (iii) 4 global (inner) detectors for trigger and centrality selection.
Neutral mesons are reconstructed through invariant mass 
analysis of decay $\gamma$ pairs detected in two types of electromagnetic calorimeters 
(15552 lead scintillator towers with 18$X_0$, and 9216 lead glass modules with 14.4$X_0$). 
The trajectories and momenta of charged hadrons in the axial %ly symmetric 
central magnetic field ($B_{max}$ = 1.15 T m) are measured by a drift chamber (DC) and 
three layers of MWPC's with pad readout (PC). Hadron identification ($\pi^\pm$, $K^\pm$, and
$p,\bar{p}$) is achieved by matching the reconstructed tracks to hits in a time-of-flight wall (TOF).

The occurrence of a p+p, d+Au or Au+Au collision (with vertex position $|z| <$ 60 cm within 
the center of the detector) is triggered by a coincidence between the two Beam-Beam Counters 
(BBC)\footnote{Plus the Zero Degree Calorimeters, ZDC, at $|\theta|<$ 2 mrad in the case of Au+Au.} 
which cover $|\eta|$ = 3.0 - 3.9. These minimum bias triggers accept respectively (52$\pm$4)\%, 
(88$\pm$4)\%, and (92$\pm$3)\% of the total inelastic cross-sections ($\sigma_{pp}^{inel}\approx$ 42 mb, 
$\sigma_{dAu}^{inel}\approx$ 2150 mb, and $\sigma_{AuAu}^{inel}\approx$ 6850 mb). 
Since hard probes are rare, high luminosities are required. PHENIX is an efficient high interaction rate 
experiment with a state-of-the-art data acquisition system capable of recording 120 MB/s to disk 
with event sizes of $\sim$100 KB, and event rates of $\sim$1-2 KHz. A total of 0.2$\cdot$10$^9$, 5.5$\cdot$10$^9$, 
and 4.0$\cdot$10$^9$ events have been sampled in Au+Au, d+Au, and p+p collisions respectively
at $\sqrt{s_{_{NN}}}$ = 200~GeV.

\section{Scaling of yields and cross-sections from p+p to A+B}
Insights on the mechanisms of particle production (and ``destruction'') in nucleus-nucleus
(A+B) collisions are obtained from the study of the scaling behavior of their yields with respect to 
p+p collisions. Depending on the $p_T$ range in which they are produced, two
different ``scaling laws'' are relevant:
%\begin{itemize}
%\item 

$\bullet$ \underline{Soft particle production} at low $p_{T}$ ($p_{T}\lesssim$ 1 GeV/$c$)
in A+B reactions is dominated by non-perturbative nucleon-nucleon collisions 
%the bulk of the A+A cross-sections is dominated by non-perturbative processes involving relatively 
with small momentum transfers. %In such conditions, energy-momentum conservation 
Since those collisions occur in ``parallel'' and with large probabilities, they 
suffer destructive interference which effectively limits the maximal number of processes 
that can lead to secondary particle production~\cite{capella}. As a result soft particle yields 
turn out to be proportional to the average number of participating  (or ``wounded'') nucleons, 
i.e. nucleons which undergo at least one inelastic collision~\cite{wnm}. 
Thus, in a A+B reaction at impact parameter $b$: 
%which are poorly localized and the involved cross-sections are large, 
%the interior nucleons are geometrically ``shadowed'' and 
\begin{equation}
E\frac{dN^{soft}_{AB}(b)}{d^3p} = \langle N_{part}(b)\rangle\;\cdot E\frac{dN_{pp}^{soft}}{d^3p},
\label{Npart_scaling}
\end{equation}
where $N_{part}(b)$ can be obtained from the nuclear thickness functions 
$T_{A,B}(b)=\int dz\, \rho_{A,B}(z,b)$ (normalized to $A,B$ resp.) 
via a Glauber model\footnote{The corresponding formulas can be found e.g. in~\cite{wnm}.}.
%\begin{equation}
%\small{
%N_{part}(b)=A\int d^2s\, T_A(s)\left[1-e^{-\sigma_{NN}T_B(s-\frac{b}{2})}\right]
%+ B\int d^2s \,T_B(s-b)\left[1-e^{-\sigma_{NN}T_A(s)}\right]}
%\label{Npart}
%\end{equation}
Such a ``participant scaling'' is indeed approximately observed in the total multiplicities
measured in heavy-ion collisions~\cite{phenix_dNdy}. Integrating $N_{part}(b)$ over impact 
parameter one obtains the average number of participant nucleons in a minimum bias A+B collision: 
$\langle N_{part}\rangle|_{{\mbox{\tiny $MB$}}}=(A\,\sigma_{pB}^{inel}+B\,\sigma_{pA}^{inel})/\sigma_{AB}^{inel}$. 
Using this expression, Eq. (\ref{Npart_scaling}), and %the fact that 
$dN^{X}_{AA}=d\sigma^{X}_{AA}/\sigma_{AA}^{inel}$, the soft invariant yields and cross-sections 
in a symmetric A+A minimum bias collision\footnote{Also, 
for min. bias A+A collisions: $\langle N_{part}\rangle|_{{\mbox{\tiny $MB$}}}=
2A\sigma_{pA}^{inel}/\sigma_{AA}^{inel}\approx 2A\pi R_A^2/\pi(2R_A)^2 \approx A/2$.} 
are related to the corresponding p+p ones via:
\begin{eqnarray}
\left.E\frac{dN^{soft}_{AA}}{d^3p}\right|_{{\mbox{\tiny $MB$}}} = \frac{2A\,\sigma_{pA}^{inel}}{\sigma_{AA}^{inel}}\cdot E\frac{dN_{pp}^{soft}}{d^3p},
\;\mbox{ and }\;
%\label{Npart_av1}\\
\left.E\frac{d\sigma^{soft}_{AA}}{d^3p}\right|_{{\mbox{\tiny $MB$}}} = \frac{2A\,\sigma_{pA}^{inel}}{\sigma_{pp}^{inel}}\cdot E\frac{d\sigma_{pp}^{soft}}{d^3p}.
\label{Npart_av2}
\end{eqnarray}
%\item 
$\bullet$ \underline{Hard particle production} at high $p_{T}$ ($p_{T}\gtrsim$ 2 GeV/$c$)
results from incoherent parton-parton scatterings with large $Q^2$. In this
regime, the pQCD ``factorization theorem''~\cite{factor} holds and the inelastic 
cross-section for the production of a given particle %in a hadronic collision
can be separated %results from the convolution 
in the product of long-distance (non-perturbative parton distribution functions, $f_{a/A}$, and 
fragmentation functions, $D_{c/h}$) and short-distance (parton-parton scattering) contributions:
\begin{equation}
E\frac{d\sigma^{hard}_{AB\rightarrow h}}{d^3p} = f_{a/A}(x,Q^2)\otimes f_{b/B}(x,Q^2)\otimes 
\frac{d\sigma_{ab\rightarrow c}^{hard}}{d^3p} \otimes D_{c/h}(z,Q^2).%+ \mathcal{O}(1/Q^2)
\label{factorization}
\end{equation}
The assumption of incoherent scattering %(and factorization) 
at high $p_T$ entails also that $f_{a/A}=A\cdot f_{a/N}$, i.e. that the density of 
partons in a nucleus $A$ should be equivalent to the superposition of $A$ 
independent nucleons. Thus,
\begin{equation}
E\frac{d\sigma^{hard}_{AB\rightarrow h}}{d^3p} = A\, B\;\cdot f_{a/p}(x,Q^2)\otimes \;f_{b/p}(x,Q^2)\otimes 
E\frac{d\sigma_{ab\rightarrow c}^{hard}}{d^3p} \otimes D_{c/h}(z,Q^2),% + \mathcal{O}(1/Q^2)
\label{factorization2}
\end{equation}
%Accordingly, %interaction amplitudes for hard processes (with small cross-sections) add incoherently 
and minimum bias hard cross-sections in A+B are expected to scale simply as
%with the number of individual scattering centers:
\begin{equation}
\left.E\frac{d\sigma^{hard}_{AA}}{d^3p}\right|_{{\mbox{\tiny $MB$}}} = A\, B\cdot E\frac{d\sigma_{pp}^{hard}}{d^3p}.
\label{A2scaling}
\end{equation}
In the general case, for a given A+B reaction with impact parameter $b$: 
\begin{equation}
E\frac{dN^{hard}_{AB}(b)}{d^3p} = \langle T_{AB}(b)\rangle \cdot E\frac{d\sigma_{pp}^{hard}}{d^3p}, 
\label{TAB_scaling}
\end{equation}
where $T_{AB}(b)=\int d^2s\,T_A(s)T_B(b-s)$ (normalized to $A\, B$) is the 
nuclear overlap function\footnote{The ``natural'' magnitudes of $T_A(b)$ and $T_{AB}(b)$ are 
$A/\pi R^{2}_{A}$ and $AB/\pi (R_A+R_B)^{2}$ respectively~\cite{eskola}.} at $b$. 
Since $T_{AB}$ is proportional to the number of 
nucleon-nucleon ($NN$) collisions: $T_{AB}(b)=N_{coll}(b)/\sigma_{pp}^{inel}$, 
one alternatively quotes Eq. (\ref{TAB_scaling}) 
in the form of ``(binary) collision scaling'' of invariant yields:
\begin{equation}
E\frac{dN^{hard}_{AB}(b)}{d^3p} = \langle N_{coll}(b)\rangle \cdot E\frac{dN_{pp}^{hard}}{d^3p}.
\label{Ncoll_scaling}
\end{equation}
From (\ref{A2scaling}) and (\ref{Ncoll_scaling}), it is easy to see that\footnote{Also, for min. bias A+A collisions:
$\langle N_{coll}\rangle|_{{\mbox{\tiny $MB$}}}=A^2\sigma_{pp}^{inel}/\sigma_{AA}^{inel}\approx
A^2\sigma_{pp}^{inel}/\pi(2R_A)^2\approx A^{4/3}/4$.} 
$\langle N_{coll}\rangle|_{{\mbox{\tiny $MB$}}}=A\, B\cdot\,\sigma_{pp}^{inel}/\sigma_{AB}^{inel}$.

Following Eqs. (\ref{Npart_av2}) and (\ref{A2scaling}), 
since $\sigma^{inel}_{pA,AA}\propto R^{2}_{A}\propto A^{2/3}$, 
the atomic number dependence of soft processes in symmetric A+A reactions is of the type
\begin{equation}
\left.E\frac{dN^{soft}_{AA}}{d^3p}\right|_{{\mbox{\tiny $MB$}}} \propto A\;\cdot E\frac{dN_{pp}^{soft}}{d^3p}\;\;\mbox{ and }\;\;
\left.E\frac{d\sigma^{soft}_{AA}}{d^3p}\right|_{{\mbox{\tiny $MB$}}} \propto A^{5/3}\;\cdot E\frac{d\sigma_{pp}^{soft}}{d^3p},
\label{A_scaling}
\end{equation}
whereas the $A$ dependence of hard processes is of the form:
\begin{equation}
\left.E\frac{dN^{hard}_{AA}}{d^3p}\right|_{{\mbox{\tiny $MB$}}} \propto A^{4/3}\cdot E\frac{dN_{pp}^{hard}}{d^3p}\;\;\mbox{ and }\;\;
\left.E\frac{d\sigma^{hard}_{AA}}{d^3p}\right|_{{\mbox{\tiny $MB$}}} = A^{2}\cdot E\frac{d\sigma_{pp}^{hard}}{d^3p}.
\label{A4_3_scaling}
\end{equation}
%\end{itemize}
\section{High $p_T$ production: Au+Au vs. p+p}
\begin{figure}[ht]
\begin{tabular}{cc}
   \includegraphics[height=5.3cm]{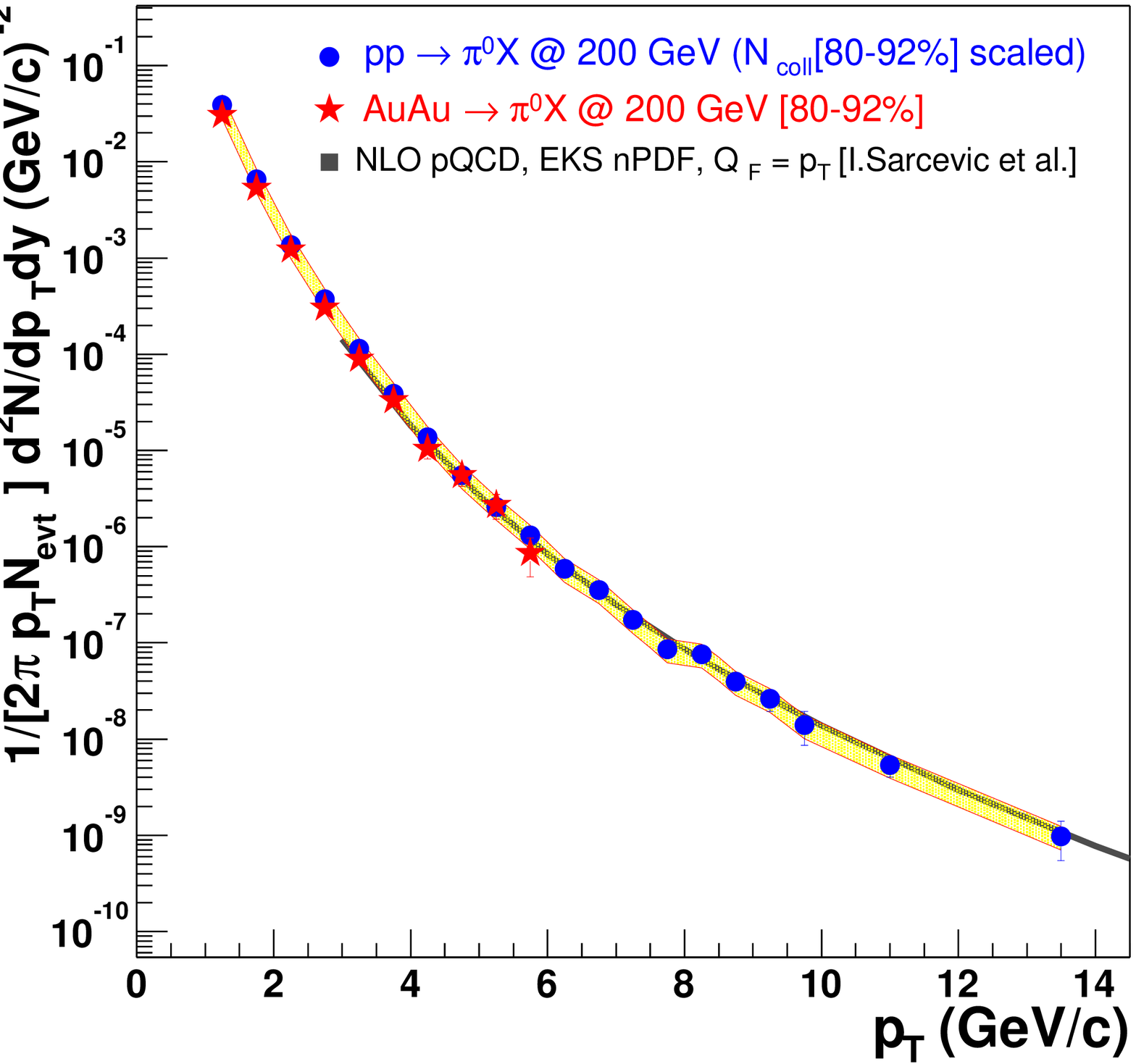} &
   \includegraphics[height=5.3cm]{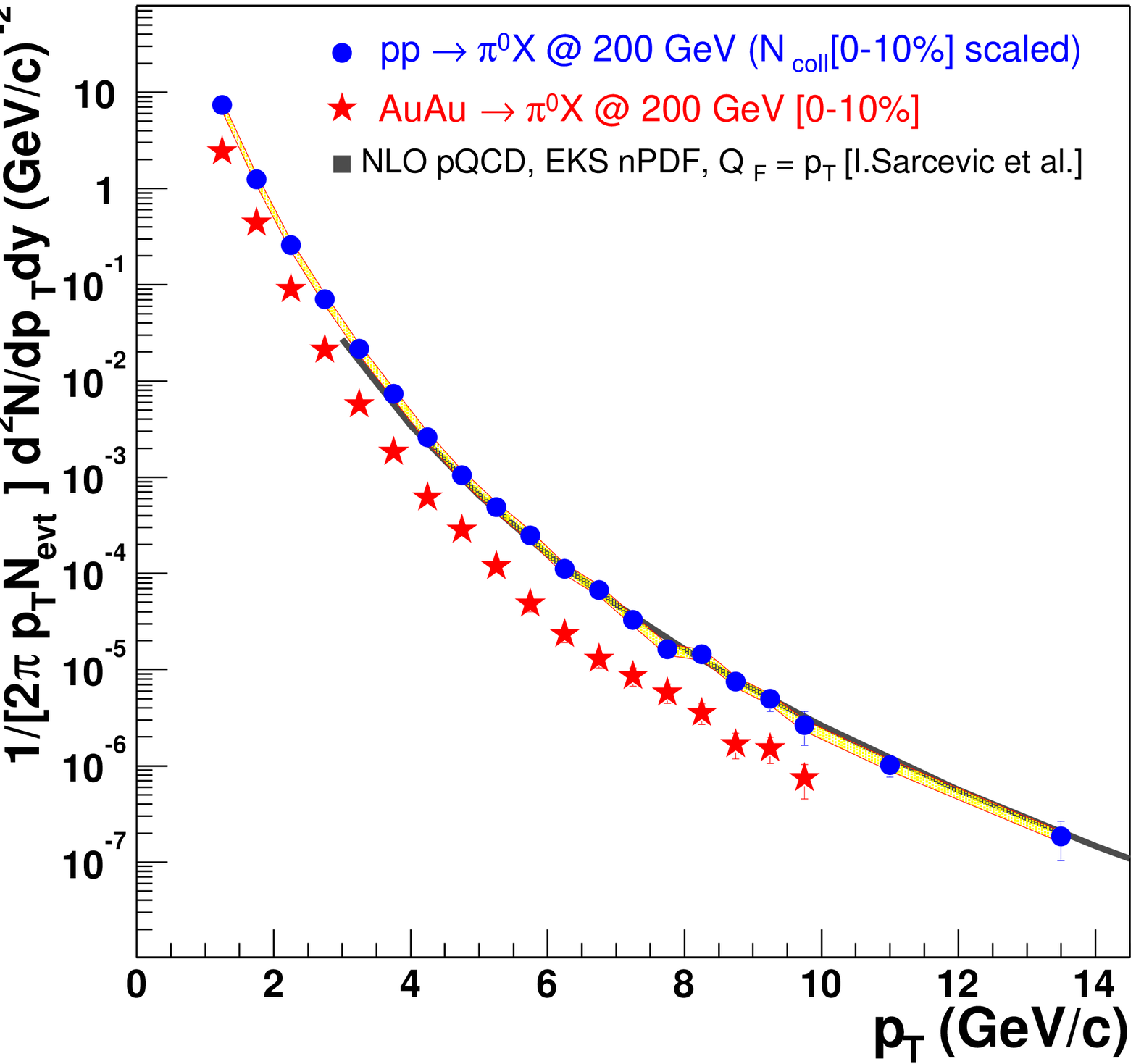} \\
\end{tabular}
%\vspace*{-0.3cm}
\caption{Invariant $\pi^0$ yields measured by PHENIX in peripheral (left) and in central (right) 
Au+Au collisions (stars)~\protect\cite{ppg014}, compared to the $T_{AB}$ scaled p+p $\pi^0$ cross-section 
(circles)~\protect\cite{phenix_pp_pi0} and to a NLO pQCD calculation (gray line)~\protect\cite{ina}. 
The yellow band around the scaled p+p points represents the overall normalization uncertainties.}
\label{fig:phenix_pi0_pp_AuAu}
\end{figure}

One of the most interesting experimental results at RHIC so far is the 
{\it breakdown} of the expected incoherent parton scattering assumption for high $p_{T}$ production in 
non-peripheral Au+Au collisions. Fig. \ref{fig:phenix_pi0_pp_AuAu} shows the comparison 
of the measured p+p $\pi^0$ spectrum~\cite{phenix_pp_pi0} to peripheral (left) and central (right) 
Au+Au spectra~\cite{ppg014}, and to standard NLO pQCD calculations~\cite{ina}. Whereas peripheral data 
is consistent with a simple superposition of individual $NN$ collisions, central %(Eq.\ref{A2scaling}), central
data shows a suppression factor of 4 -- 5 with respect to this expectation. 
%\hspace{-1cm}

\begin{figure}[htb]
\includegraphics[height=6.cm]{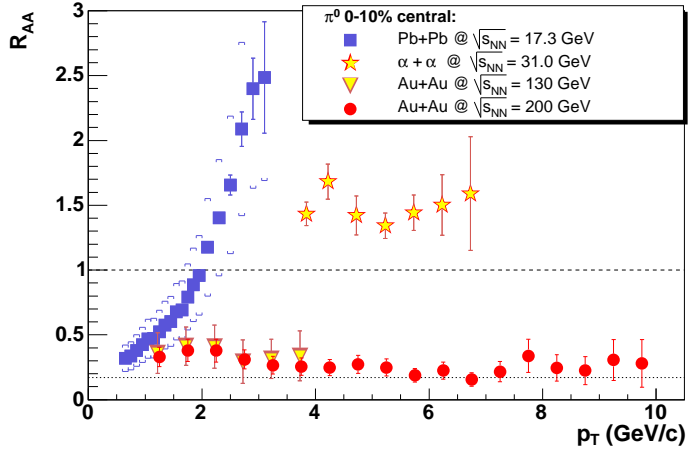}
\caption[]{Nuclear modification factor, $R_{AA}(p_T)$, for $\pi^0$ measured in
central ion-ion reactions at CERN-SPS~\cite{wa98_pi0}, CERN-ISR~\cite{ISR_pi0}, 
and BNL-RHIC~\cite{ppg003,ppg014} energies. The dashed (dotted) line is the expectation of 
``$N_{coll}$ ($N_{part}$) scaling'' for ``hard'' (``soft'') particle production.}
\label{fig:R_AA_pi0_syst}
\end{figure}

From Eq. (\ref{TAB_scaling}), it is customary to quantify the medium effects at 
high $p_{T}$ using the {\it nuclear modification factor} given by the ratio of the 
A+A invariant yields to the p+p cross-sections scaled by the nuclear overlap at impact parameter $b$: 
\begin{equation} 
R_{AA}(p_{T},b)\,=\,\frac{d^2N^{\pi^0}_{AA}(b)/dy dp_{T}}{\langle T_{AB}(b)\rangle\,\cdot\, d^2\sigma^{\pi^0}_{pp}/dy dp_{T}}.
\label{eq:R_AA}
\end{equation}
$R_{AA}(p_T)$ measures the deviation of A+A from an incoherent superposition 
of $NN$ collisions in terms of suppression ($R_{AA}<$1) or enhancement ($R_{AA}>$1). 
Figure~\ref{fig:R_AA_pi0_syst} shows $R_{AA}$ as a function of $p_T$  for several $\pi^0$ measurements 
in high-energy A+A collisions. Much of the excitement at RHIC comes from the fact that the 
PHENIX $R_{AA}$ values for central collisions at 200 GeV (circles) and 130 GeV (triangles) 
are noticeably below unity in contrast to the enhanced production ($R_{AA}>$1) 
observed at CERN-ISR~\cite{ISR_pi0} (stars) and CERN-SPS~\cite{wa98_pi0} 
(squares) energies. This enhanced production, observed first in p+A fixed-target 
experiments~\cite{cronin} (``Cronin effect''), is interpreted in terms of multiple initial-state 
soft and semi-hard interactions which broaden the transverse momentum of the colliding 
partons prior to the hard scattering itself.

%\hspace{1cm}
The breakdown of the expectations from collinear factorization for high $p_T$ production
in central A+A collisions at RHIC by such a large factor, has been interpreted as due 
%could be due in principle  
to one (or more) of the following facts:
\begin{enumerate}
\item Breakdown of leading-twist QCD collinear factorization itself:
The incoherence between long- and short-distance effects in which the
factorized product Eq. (\ref{factorization}) relies upon, does not hold for A+A collisions.
\item Strong {\bf initial-state} effects: The parton distribution functions in the nuclei are 
strongly modified: $f_{a/A}<<A\,\cdot f_{a/p}$ in the relevant ($x,Q^2$) range, resulting 
in a reduced number of effective partonic scattering centers in the initial-state.
\item Strong {\bf final-state} effects: The parton fragmentation functions (or, more generally, 
any post hard collision effect on the scattered partons) are strongly modified in the 
nuclear medium compared to free space.
\end{enumerate}
Explanations 1. and 2. are usually invoked in the context of the ``Color-Glass-Condensate''
picture~\cite{CGC} which assumes that the kinematical conditions prevailing in the initial-state of 
an atomic nucleus boosted to RHIC energies (moderate $Q^2\approx Q^2_s$ = 1--2 GeV$^2/c^2$, 
and small parton fractional momenta $x\lesssim 10^{-2}$) are such that nonlinear QCD effects 
($g+g\rightarrow g$ processes, amplified by a $A^{1/3}$ factor compared to the proton case) 
are important and lead to a strong saturation of the parton (mostly gluon) densities in the nuclei. 
In this scenario, one expects $N_{part}$ (instead of $N_{coll}$) scaling at moderately high $p_T$'s~\cite{dima},
as approximately observed in the data (dotted line in Fig.~\ref{fig:R_AA_pi0_syst}).
Explanation 3., on the other hand, relies on the expectations of ``jet quenching'' in a Quark 
Gluon Plasma~\cite{qgp_en_loss} in which the hard scattered partons lose energy by final-state 
``gluonstrahlung'' in the dense partonic system formed in the reaction. After traversing the medium,
the partons fragment into high $p_T$ (leading) hadrons with a reduced energy compared to standard 
fragmentation in the ``vacuum''. Different jet quenching calculations %~\cite{vitev,xnwang,arleo,levai_pp_pi0,urs} 
can reproduce the magnitude of the $\pi^0$ suppression assuming the formation of a hot 
and dense partonic system characterized by different, but closely related, properties: 
i) large initial gluon densities $dN^{g}/dy\approx$ 1000~\cite{vitev}, ii) large ``transport coefficients'' 
$\hat{q}_{0}\approx$ 3.5  GeV/fm$^2$~\cite{arleo}, iii) high opacities $L/\lambda\approx$ 3.5~\cite{levai_pp_pi0}, 
iv) effective parton energy losses of the order of $dE/dx\approx$ 14 GeV/fm~\cite{xnwang}, or 
v) plasma temperatures of $T\approx$ 0.4 GeV~\cite{moore}.
%It turns out, however, that the centrality, $p_T$, and $\sqrt{s}$ dependence 
%of the high $p_T$ suppression in Au+Au collisions~\cite{summ_dde} cannot solely distinguish 
%between both scenarios and, thus, this motivates the investigation of a system like d+Au at RHIC, 
%where final-state medium effects are absent.

%The observed high $p_T$ suppression has been interpreted in terms of two different
%QCD scenarios:
%\begin{enumerate}
%\item {\bf Final-state} effects in a Quark-Gluon-Plasma: The magnitude and $p_T$ dependence 
%of $R_{AA}$ is consistent with the predictions of ``jet quenching'' models~\cite{qgp_en_loss} 
%in which the hard scattered partons lose energy by final-state ``gluonstrahlung'' in the dense and 
%hot partonic system formed in the reaction. After traversing the medium, the partons fragment
%into high $p_T$ (leading) hadrons with a reduced energy. Very high gluon densities 
%($dN^g/dy\sim$ 1000~\cite{vitev}) are required in order to reproduce the observed 
%quenching of the high $p_T$ spectra. 
%\item {\bf Initial-state} effects in a highly dense and saturated Au parton wave-function: 
%\end{enumerate}
%in the small x region one can probe nonlinear QCD effects and explore the fundamental question of 
%what kind of dynamics slows down and eventually stops the rapid growth of the cross section 
%(or the structure functions F2(x) and xG(x)) at small x. The use of nuclei allows one to enhance these effects.
%Constraints from unitarity [107, 109] indicate that the growth of the parton densities in nuclei 
%should be tamed at significantly larger x than in the proton.

\section{High $p_T$ production: d+Au vs p+p}
\begin{figure}[htb]
\includegraphics[height=5.9cm]{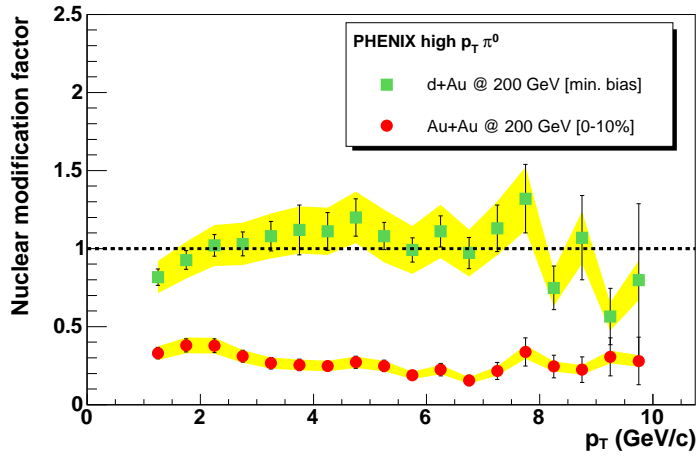}
\caption[]{Nuclear modification factor, $R_{AA}(p_T)$, for $\pi^0$ measured at $y$ = 0 in
minimum bias d+Au (squares)~\cite{ppg028} and central Au+Au (circles)~\cite{ppg014} reactions at  
$\sqrt{s_{_{NN}}}$ = 200~GeV.}
\label{fig:R_dAu_AuAu_pi0}
\end{figure}

In order to disentangle between the two different (QGP and CGC) QCD scenarios, it was of paramount 
interest to determine experimentally the modification, if any, of high $p_T$ hadron production 
due to initial-state nuclear effects alone, i.e. for a system in which a hot and dense medium is 
not produced in the final state. The results of high $p_T$ $\pi^0$ production at midrapidity in d+Au 
collisions at $\sqrt{s_{_{NN}}}$ = 200~GeV~\cite{ppg028} do not show any indication of suppression (Fig.
\ref{fig:R_dAu_AuAu_pi0}). On the contrary, $\pi^0$ production seems to be slightly enhanced 
($R_{dAu}\approx$ 1.1) compared to the expectations of collinear factorization.
%\hspace{1cm}

This result, reminiscent of the ``Cronin enhancement'', indicates that the observed 
suppression in Au+Au central collisions at mid-rapidity is not an initial-state effect arising 
from strong modifications of the gluon distribution functions in nuclei at moderately small
values of parton fractional momenta $x$, but more likely a final-state effect of the produced dense medium.

\section{Summary}
PHENIX has measured $\pi^0$ at mid-rapidity up to $p_T\approx$ 10~GeV$/c$ 
in p+p, d+Au and Au+Au collisions at $\sqrt{s_{_{NN}}}$ = 200~GeV. 
The spectral shape and  invariant yields in peripheral Au+Au reactions are consistent with 
those of p+p reactions scaled by the number of inelastic $NN$ collisions, in agreement with 
pQCD collinear factorization expectations. In central Au+Au reactions, the $\pi^0$ are suppressed 
by a factor 4--5 with respect to the same expectations. The magnitude of this deficit can be 
reproduced by pQCD-based parton final-state energy loss calculations in an opaque medium, 
as well as by initial-state gluon saturation models. The unquenched high $p_T$ $\pi^0$ production 
in minimum bias d+Au collisions, however, seems to rule out any significant ``cold'' nuclear 
matter effect as responsible for the observed suppression in Au+Au central reactions.

%% Bibliography made with BibTeX:
%%Use use the bibliographic style 'amsunsrt.bst'
%\bibliographystyle{amsunsrt}
%\chapbblname{sample} % \chapbblname{<name of .bbl file>}
%\chapbibliography{sample} % \chapbibliography{ <name of .bib file>}

\end{document}